\documentclass[11pt,twoside]{article}
\usepackage{asp2010}
\usepackage{natbib}

\resetcounters

\begin{document}

\title{Can Li-rich K giants eject shells? Assembling the lithium puzzle 
in K giants} 
\author{R.~de~la~Reza$^1$, N.A.~Drake$^{1,2}$
\affil{$^1$Observat\'orio Nacional/MCT, Rua Gal. Jos\'e Cristino 77,
Rio de Janeiro CEP 20921-400, Brazil}
\affil{$^2$Sobolev Astronomical Institute, Saint Petersburg State University, 
Saint Petersburg 198504, Russia}
}

\begin{abstract}
The existence of K giant stars with high Li abundance continues to challenge 
the standard theory of stellar evolution. 
All recent extensive surveys in the Galaxy show the same result:
about 1\% of the mainly normal slow rotating K giants are Li rich. 
We explore here a model with two scenarios based on the important relation 
of Li-rich and Li-poor K giants with IR excesses. 
In this model, all K giant stars suffer a rapid enrichment and
depletion of Li inducing the formation and ejection of circumstellar shells. 
The observational detection of these shells will not only validate this model, 
but also will give important hints on the mechanism of Li enrichment of these 
stars. 
\end{abstract}

\section{The lithium puzzle}

Following the standard stellar evolution theory, Li-rich K giants don't exist. 
However, they do exist and the failure to completely explain this situation is 
known in the literature as ``the puzzle of lithium-rich K giant stars''. 
In fact, this fragile element ($^7$Li) can survive only in the low temperature 
surface layers of stars. 
As the stars evolve in the red giant branch (RGB), the reminiscent main sequence 
$^7$Li abundance is reduced by the action of the first dredge-up.
Standard models predict that $^7$Li is depleted by a factor of 60 in low mass, 
solar metallicity stars (Iben 1967). 
An initial meteoritic Li abundance, $\log\varepsilon {\rm (Li)}\sim 3.2$
(where $\log\varepsilon{\rm (Li)} =\log (N_{\rm Li}/N_{\rm H}) + 12.0$), will 
be reduced to about 1.5 in the RGB stars. 
In reality, RGB stars have been found even more 
depleted in lithium than this predicted value. 

Wallerstein \& Sneden (1982) discovered, unexpectedly, the first K giant star 
with a large Li abundance. 
After the discovery of some additional Li-rich K giants, a first survey among bright,  
nearby K giants by Brown et al. (1989), showed that about 1-2~\% of K giants were Li rich. 
At that epoch, nearly 20 Li K giants were already known. 
During the Pico dos Dias Survey (PDS), based on IRAS sources and intended 
to discover Post T-Tauri stars, a serendipitous discovery of further $\sim\! 20$ 
new Li K giants was made. 
These new objects were found among stars of a more distant population and all stars 
were IRAS sources (Greg\'orio-Hetem et al. 1992; Torres et al. 1995). 
This discovery established a real connection between Li-rich and Li-poor K giants 
and far-infrared (FIR) excesses (de la Reza et al. 1996, 1997; Castilho et al. 1995). 
In general, Li K giants are considered those stars with $^7$Li abundances higher 
than $\log\varepsilon {\rm (Li)} = 1.5$. 
Extremely high Li abundances have been found for some giants, as are the cases of 
IRAS~13539-4153 with $\log\varepsilon{\rm (Li)}= 4.2$ (Reddy \& Lambert 2005), 
PDS~68 ($\log\varepsilon{\rm (Li)}=3.9$,  Drake 1998), 
HD~19745 ($\log\varepsilon{\rm (Li)^{\rm NLTE}}=4.7$,  
$\log\varepsilon{\rm (Li)^{\rm LTE}}=4.08$, de la Reza \& da Silva 1995,  
$\log\varepsilon{\rm (Li)^{\rm LTE}}=3.90\pm 0.30$, Reddy \& Lambert 2005). 
All known Li K giants are found near the RGB bump or near the RGB tip, 
and even at the giant clump in  case of the more massive stars.

Today, up to the year of 2011, various and large surveys in the Galaxy 
have been  made searching for new Li K giants. 
In Table~1 all these surveys are described. The results of these surveys are 
important because different regions of the Galaxy as the thin and thick disks 
and the bulge were explored.  
Also, these surveys covered giant stars of different metallicities (solar and 
metal deficient) and of different stellar masses.

\begin{figure}[htbp]
 \plottwo{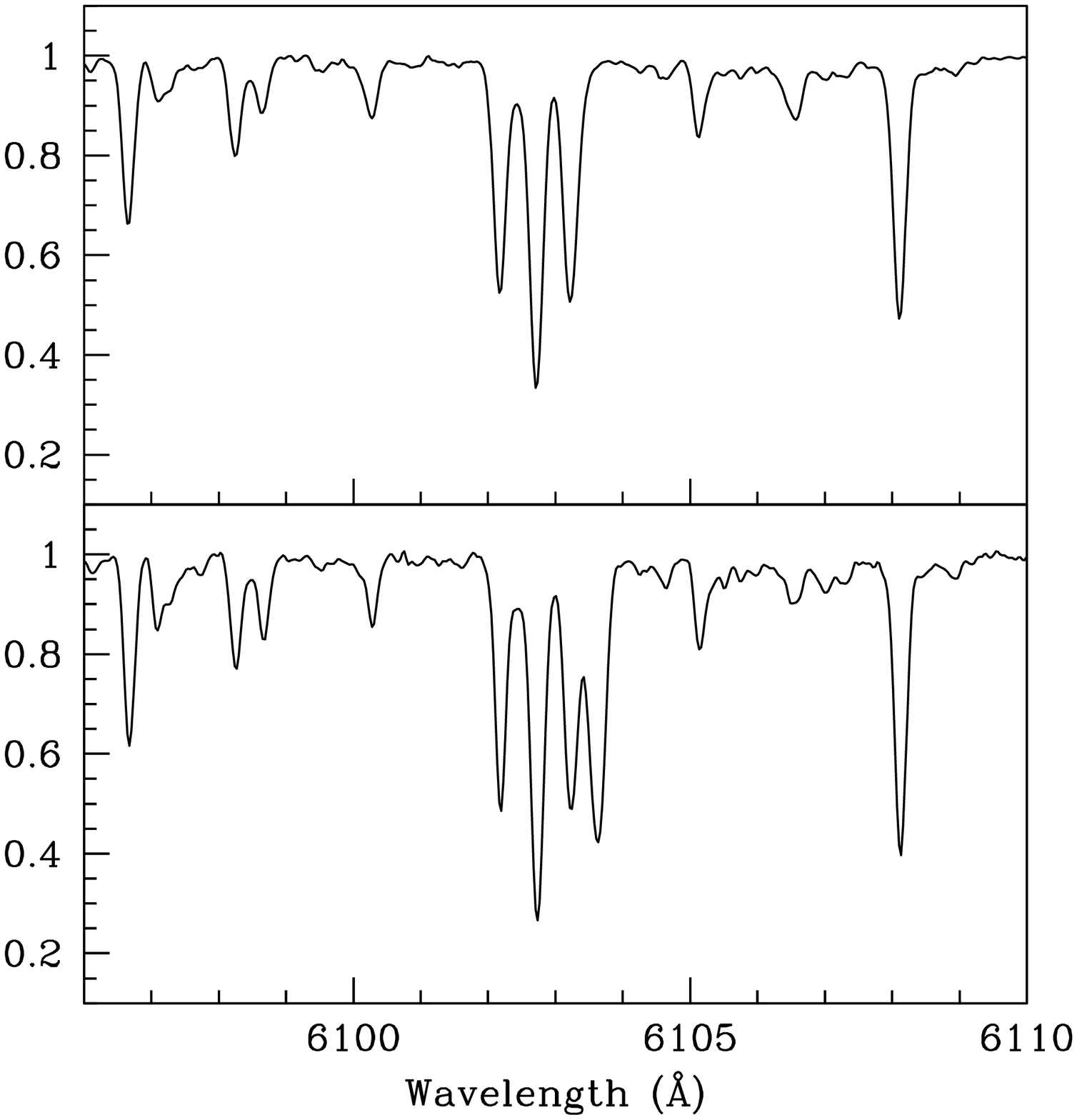}{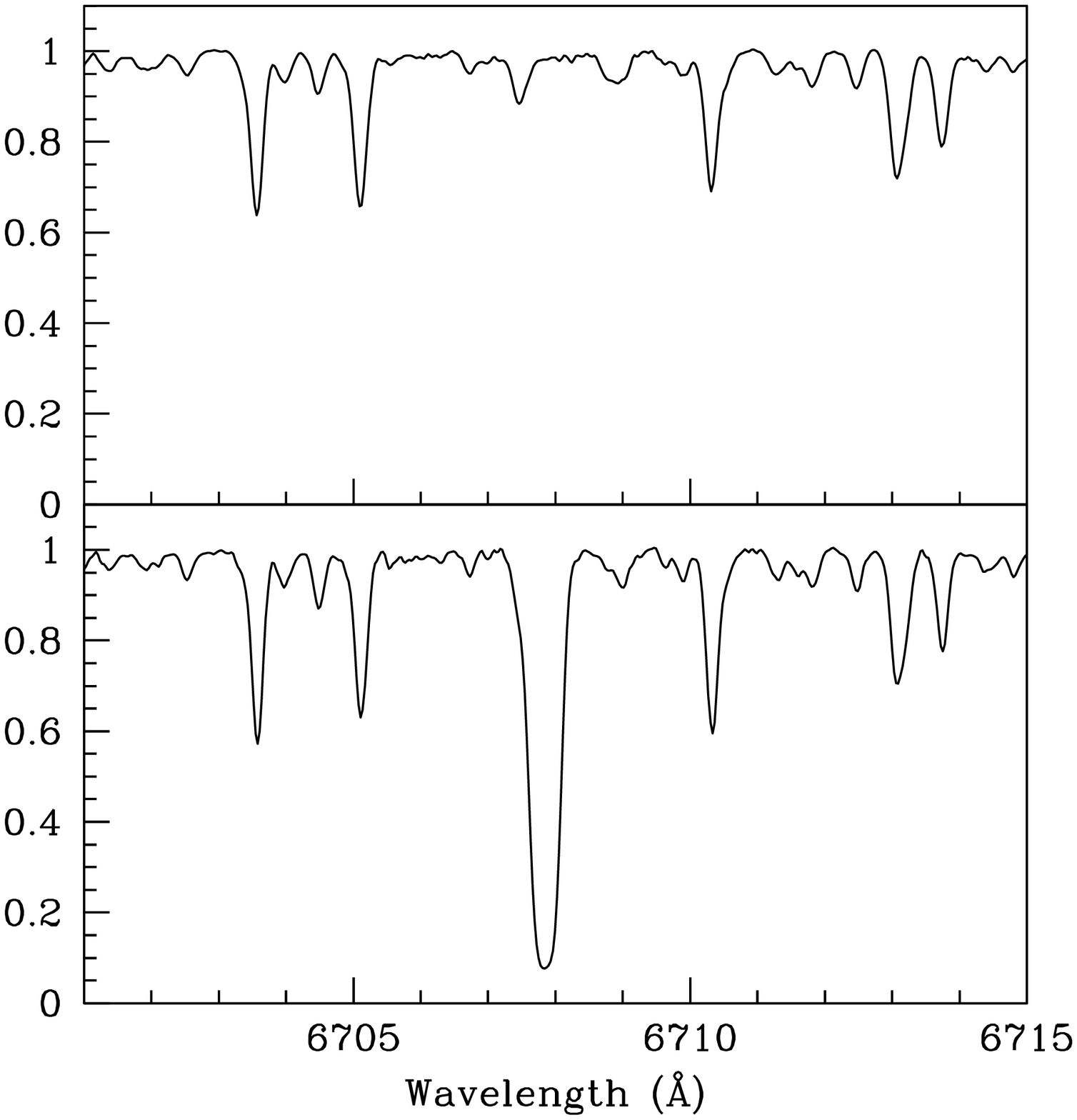}
\caption{Example of the spectra of normal Li-poor giant and Li-rich K giant 
in the regions of the \ion{Li}{i} lines at 6104~\AA\ and 6708~\AA. 
{\it Upper panel} - normal K giant, {\it bottom panel} - Li-rich K giant.}  
 \end{figure}

In all these cases, and quite surprisingly in a certain way, the rate of Li K 
giants is always around 1~\%. There is, however, a noticeable exception: 
among the less numerous fast rotating K giants (with 
$v\sin i \ge  8$~km\,s$^{-1}$) the rate of Li K giants increases to 
$\sim\! 50$~\% ! (Drake et al. 2002). 
We must note that in this case almost all of these fast rotating stars presented 
FIR excesses. We can conclude that the main surveys described in Table~1 with 
a rate of Li K giants of about 1\% represent more the normal, slow rotating 
($v\sin i < 8$~km\,s$^{-1}$) K giants population. 
It is important to note that the Li-rich and Li-poor K giants have the same general 
stellar properties, being  the $^7$Li abundance the sole difference. 
This can be seen, for instance, in the spectra of both, Li-poor and Li-rich, 
K giants shown in Figure~1.

\begin{table*}[h]
\label{t1}
\begin{center}
\caption{Surveys for Li-rich K giants}
\begin{tabular}{ccccl}
\hline
Candidates &  Properties  & Li giants  & Rate & $\;\;\;\;\;\;$References   \\
\hline
670        & Nearby objects  &      10       & 1.5 \% & Brown et al.  (1989) \\
           & IRAS (distant objects) &  19 &         & de la Reza et al. (1997) \\
280        & Selected IRAS   &  5            & 1.7 \% & Castilho et al.  (1998) \\
     &  High rotating giants &               & 50 \%  & Drake et al. (2002)  \\
400        &  Galactic Bulge &  2            & 0.5 \% & Gonzalez et al. (2009)  \\
401        &  Galactic Bulge &  3            & 0.7 \% & Lebzelter et al. (2011) \\
824    & Galactic thick disk &  6            & 0.7 \% & Monaco et al. (2011) \\
2000       & Extended survey & 19            & 1.0 \% & Kumar et al. (2011) \\
700        & Metal poor      &  6            & 0.8 \% & Ruchti et al. (2011) \\
\hline
\end{tabular}
\end{center}
\end{table*}

\section{The lithium enrichment models}

Some tentative models were published trying to explain the Li K giant phenomena.
Some of these models invoke pure  external, other pure internal causes or a mixture of both.
Those with  external ones propose that the giant star has engulfed a planet or a brown dwarf. 
This could explain, in principle, the increasing Li abundance in the star resulting 
from the swallowing an external object and explaining at the same time an 
increase of the stellar rotation by transfer of momentum (Alexander 1967; 
Siess \& Livio 1999).  
The pure internal model (Palacios et al. 2001) is based on an extra mixing 
parameter, also known as a diffusion coefficient $D_{\rm mix}$.  
A canonical value of $D_{\rm mix}$ does not conduce to the production of 
extra $^7$Li which requires an increase of $D_{\rm mix}$ by a factor of 100.
However, the increasing of  $D_{\rm mix}$ is for the moment an {\it ad-hoc} 
hypothesis and the mechanism needs a self-consistent treatment.  
Denissenkov \& Weiss (2000) and Denissenkov \& Herwig (2004) proposed a mixed 
scenario in which the engulfment 
of a planet provokes the required increase of $D_{\rm mix}$ producing the 
extra-mixing and leading to the increase of $^7$Li in the stellar atmosphere. 
Note that following these authors a tidal effect in a binary star could also 
produce this effect. 

Nevertheless, the ``engulfing'' models, pure or mixed, have severe problems due 
to the following considerations: 
{\it a)}  these  models may be applied mainly to rapid rotating 
giants while the majority of Li-rich giants are slow rotators; 
{\it b)} the engulfing process can happen everywhere in the RGB and this is 
not observed, because Li K giants are concentrated mostly close to the RGB bump;
{\it  c)} finally, enrichment with the element $^9$Be is expected together with 
$^7$Li in the planet engulfing scenario which  is not observed in the Li K giants 
(Melo et al. 2005). 

In the ``Lithium flash'' model proposed by Palacios et al. (2001), a thermal 
instability causes the $^7$Li flash increasing the Li abundance with a 
duration of $\sim\! 10^4$~yr. 
This produces an increase of the stellar luminosity thus explaining the mass loss. 
We also note that a rapidly rotating single G8~II giant, with high Li abundance 
and a magnetic field, have been detected by L\`ebre et al. (2009). 
As far as we know, this is the only giant known with these properties.

\begin{figure}[t]
 \plottwo{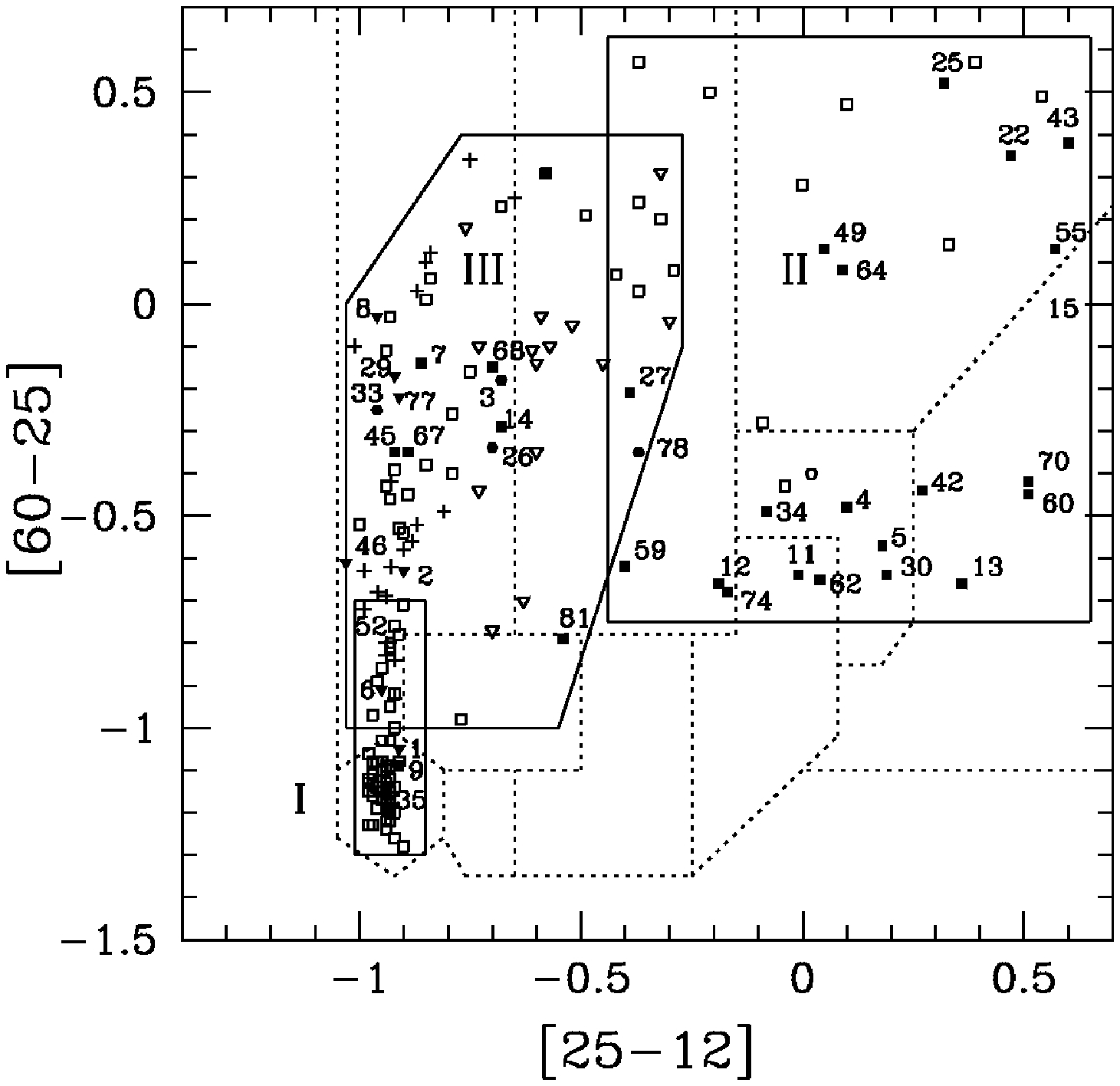}{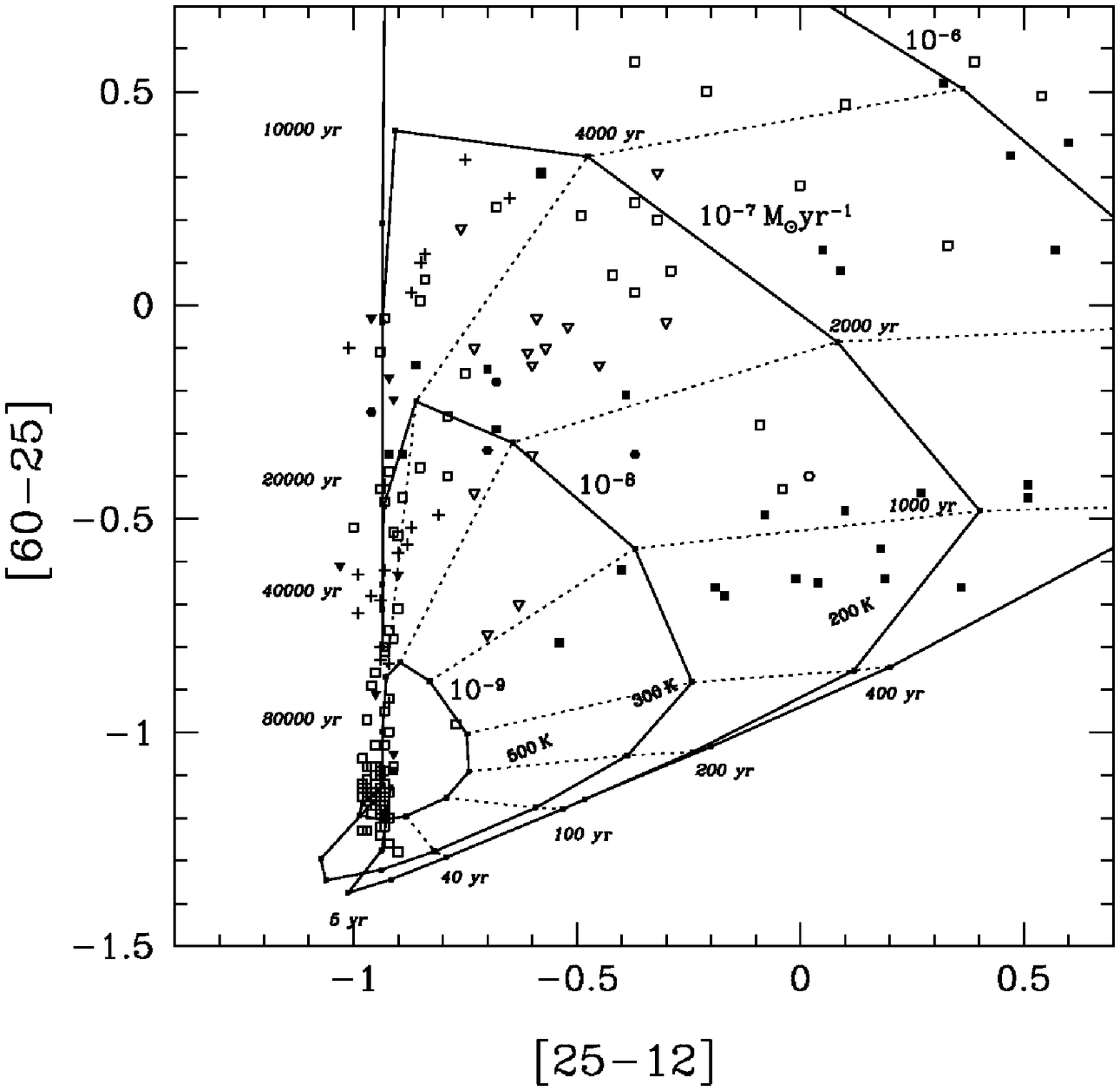}
\caption{{\it Left (a):} Distribution of IRAS sources corresponding to K giants 
contained in three regions labeled I, II, and III. 
The regions marked by broken lines are those defined by van der Veen \& 
Habing (1988). 
Labeled filled symbols correspond to Li-strong K giants: {\it squares} 
(three fluxes are of good quality), {\it triangles} (one flux is only an upper 
limit), and {\it hexagons} (two fluxes are only upper limits). 
Corresponding open symbols represent Li-weak K giants. 
 {\it Right (b):} Same points are presented together with evolution curves of CS. 
The four curves calculated for an CS expansion velocity of 2~km\,s$^{-1}$, 
star temperatures and radii equal to 4000~K and 20~$R_\odot$ respectively, 
correspond to mass losses between $10^{-9}$ and $10^{-6}$~$M_\odot$\,yr$^{-1}$.
The 
time steps are also indicated. (Figure taken from de la Reza et al. 1997.)}  
 \end{figure}

\section{The lithium -- mass loss scenario}

We proposed (de la Reza et al. 1996, 1997) a model with the aim to interpret the
``high Li abundance -- FIR excess'' relation. 
Here, all K giants pass by a rapid internal enrichment of the atmosphere by 
new $^7$Li following, probably, the known ``$^7$Be mechanism''. 
This powerful event produces the formation of a spherical circumstellar 
shell, which is subsequently detached from the star. 
This happens at the RGB bump or between the bump and the tip in the RGB, 
where the majority of Li K giants is found. 
The ejected shell moves away from the star with constant (and slow) velocity.
Depending on the duration of the process of shell formation, 
we obtain two relatively different scenarios. 
In the first one, as described in de la Reza et al. (1996, 1997), 
the process is very rapid. 
Consider the FIR diagram based on  IRAS 12, 25 and 60~$\mu$m colors 
presented in Figures~2a and 2b. 
Here the star begins its sudden Li enrichment in the box~I (at the left 
bottom part of the figure) where the giants present no FIR excesses. 
The formed shell is rapidly ejected with the velocity $V = R/t$, 
where $t$ is the time and $R$ is the distance of the 
shell to the star. 
The path of the detached shells in the IRAS color-color diagram forms 
a loop (see Figure~2b) returning after to the same initial box. 
The size of the loops depends on the considered mass loss and for values of $V$, 
of the order of 1 to 2~km\,s$^{-1}$, the complete loop is realized in $\sim\! 100\,000$~yr. 
The positions of the different  Li-rich giants (black squares) or Li-poor giants 
(open squares) describe, in this scenario, the present evolutionary stage of the shell. 
It is proposed in this scenario that the Li-poor giants,  with or without 
FIR excesses, rapidly deplete the new Li. 
For this scenario, all shells are detached.

In de la Reza et al. (1996, 1997) no mention is made about the
mechanism of Li enrichment. 
Probably, the most interesting model to be considered for this case is 
``The lithium flash: thermal instabilities generated by lithium 
burning in RGB stars'' as proposed by Palacios et al. (2001). 
In any case, the first scenario of de la Reza et al. (1996, 1997) 
has the disadvantage of requiring an extremely 
rapid, almost explosive, $^7$Li enrichment phenomenon. 
Also, there is no way of predicting the size of the shell. 

The second scenario, equally physically compatible with the shell 
evolution as presented in the Figure~2b, can be presented.
Here, the $^7$Li enrichment mechanism is slower than in the first 
scenario and could last some few thousand years. As before, the giant 
star begins its enrichment in the same, non-FIR excess, box mentioned above, 
but differently, an attached, gradually increasing, shell is formed, 
with the same velocity $V= R/t$,  $R$ being  the size of the increasing shell. 
This shell formation lasts up to the moment when the star ceases to be 
Li enriched, then the shell is detached and continues to expand with the 
same velocity,  $R$ being now the size of the shell plus the distance to the star. 
Contrary to the first scenario, the size of the shell can be estimated and 
perhaps we are in a more plausible physical situation, not requiring 
an explosive Li enrichment process. 
In this scenario it is expected that all Li-rich K giants in the IRAS diagram 
have non-detached shells, whereas the shells are detached in the case of the 
Li-poor giants.

\section{Conclusions} 

We presented here a general picture of what is called in the literature 
``the Li puzzle of K giant stars''. 
The complete collection of Galactic surveys for new Li-rich giants until 2011, 
presents the same impressive result indicating that about 1~\% of the giants are Li-rich stars. 
Being these Li K giants similar, in all their general stellar properties, to the 
Li-poor K giants, it is much probable that the Li-rich state results from a 
short episode of Li enrichment during the evolutionary life of all K giants in the RGB. 
An important exception to the frequency of Li-rich stars among K giants 
was found for the less numerous fast rotating ($v\sin i \ge 8$~km\,s$^{-1}$)
K giants, where about 50~\% are Li rich (Drake et al. 2002). 
In this case,  a fast rotating K giant, after gaining its fresh $^7$Li, 
probably prevents the $^7$Li to be depleted by the action of the stellar 
rotation. We note that an important step in the study of the properties of 
Li-rich and Li-poor giants was  the discovery of their strong relation with FIR (IRAS) excesses. 

We have outlined here two scenarios, related to the fast Li enrichment, 
depletion and mass loss,  presented originally in de la Reza et al. (1996, 1997). 
In one scenario, the very fast $^7$Li enrichment is followed by the detachment 
of a circumstellar shell. 
In the second scenario, practically the same processes occur, but much more slowly. 
Here the giant star  becoming Li rich, develops a progressively extended 
attached shell. 
Only when the internal $^7$Li enrichment ceases and the new $^7$Li is depleted, 
the shell is detached. On the other hand, we should mention that 
Pereyra et al. (2006) has performed  polarimetric measurements of Li K giants. 
They show the existence of a correlation between the  intrinsic polarization 
and the IRAS 25~$\mu$m flux, suggesting that dust scattering is the source of 
polarization and indicating the existence of non-symmetrical shells.

We believe that it will be very important to visualize these shells by imaging.  
Whether they will be detached or non-detached, will confirm one of the 
``Li -- mass loss'' scenarios and the probably non-spherical appearance of the shells.
According to the first scenario all shells must be detached, while in the second one
the shells will be attached for Li K giants and  detached for Li-poor K giants.
 In any case, these eventual shells detections will give very important hints 
to the study of the stellar $^7$Li enrichment. We must not forget that this 
mechanism represents a clear physical way of how K giants contribute, 
by means of Li-rich mass loss, to the enrichment of $^7$Li in the interstellar 
medium of the Galaxy. 

\bigskip
{\it Note added in proofs}

A recent important related work appeared in Kirby et al. (2012)
(astro.ph/1205.1057, ApJ 752, L16, 2012) referring to the discovery of 
14 Li-rich giants among 1764 giants belonging to eight dwarf spheroidal low-metal
galaxies, representing a rate of 0.85~\%.
These results include the most metal-poor Li-rich known giant 
($\log\varepsilon{\rm (Li)}^{\rm NLTE}=3.15$, [Fe/H]=--2.82). 
Because the abundance of Li is the only difference among giant stars, these authors ``consider the possibility that Li enrichment is a universal phase of evolution that affect all stars, and it seems rare only because it is brief''.

\section{Discussion at the Meeting}

{\bf A. Miroshnichenko:}  It is interesting to hear about this group of Li-rich 
K-type giants in connection with B[e] objects that were formerly called 
unclassified and which I call FS CMa stars. About 30 \% of CMa stars show 
Li lines attributed to K-type secondaries. 
It would be interesting to study the two groups together to reveal possible similarities.

{\bf R. de la Reza:} In fact, some authors as Denissenkov et al. (2004) 
have proposed that in a binary system, in which one of the members is 
a Li-rich K giant, a tidal effect  induced, in some way, the internal 
extra mixing required to achieve the Li enrichment  in the K-type secondary star.

{\bf T. Rivinius:} What ESO instrumentation/telescopes you can use to observe these shells?

{\bf R. de la Reza:}  We can use different possibilities depending on 
the distances of the targets. Maybe the best results can be obtained 
using VLTI with their main detectors, in the case of “young” shells.  
For the case of “older” detached shells ALMA could be a solution. 
               
\end{document}